\documentclass[floats,floatfix,amssymb,prd,twocolumn,superscriptaddress,nofootinbib]{revtex4-1}
	
\usepackage{subcaption}
\usepackage{ragged2e}
\DeclareCaptionJustification{justified}{\justifying}
\makeatletter
\newcommand{\subsetsim}{\mathrel{\mathpalette\subset@sim\relax}}
\newcommand{\subset@sim}[2]{%
  \vtop{\offinterlineskip\m@th
    \ialign{\hfil##\cr
      $#1\subset$\cr\noalign{\kern0.5pt}\scalebox{0.9}{$#1\sim$}\cr
    }%
  }%
}
\makeatother

\usepackage{amssymb,amsmath,verbatim,mathtools,needspace,enumitem,etoolbox,graphicx,physics,microtype,afterpage,bm}
\usepackage[dvipsnames, usenames]{xcolor}
\definecolor{linkcolor}{rgb}{0.0,0.3,0.5}
\usepackage{booktabs}
\usepackage[unicode, colorlinks=true, linkcolor=linkcolor, citecolor=linkcolor, filecolor=linkcolor,urlcolor=linkcolor, pdfusetitle]{hyperref}
\usepackage[all]{hypcap}
\usepackage[T1]{fontenc}
\usepackage[utf8]{inputenc}
\usepackage{tabularx}
\usepackage{float}
\usepackage{empheq}

\newcommand{\unmezzo}{\frac{1}{2}}
\newcommand{\F}{{\rm F}}
\newcommand{\B}{{\rm B}}

\usepackage{multirow}
\usepackage{pifont}
\usepackage{lmodern}

\usepackage{multirow}

\allowdisplaybreaks
\usepackage{tikz}
\usepackage{color}
\usepackage{framed}
\usepackage{hyperref}
\hypersetup{colorlinks, citecolor=bluscuro, linkcolor=black, urlcolor=bluscuro}
\definecolor{rossos}{cmyk}{0,1,1,0.55}
\definecolor{bluscuro}{rgb}{0.15, 0.2, .85}
\definecolor{bluchiaro}{cmyk}{1,.3,0.,0.1}
\definecolor{ForestGreen}{rgb}{0.13, 0.55, 0.13}
\definecolor{azure}{rgb}{0.0, 0.5, 1.0}
 
\usepackage{slashed}

\def\nn{\nonumber}

\def\bea{\begin{eqnarray}}
\def\eea{\end{eqnarray}}

\def\d{{\mathrm{d}}}

\newcommand{\bs}{\begin{subequations}}
\newcommand{\es}{\end{subequations}}

\newcommand{\be}{\begin{equation}}
\newcommand{\ee}{\end{equation}}
\renewcommand{\d}{{\rm d}}

\newcommand{\lp}{\left (}
\newcommand{\rp}{\right )}

\def\lsim{\mathrel{\rlap{\lower4pt\hbox{\hskip0.5pt$\sim$}}
    \raise1pt\hbox{$<$}}}         
\def\gsim{\mathrel{\rlap{\lower4pt\hbox{\hskip0.5pt$\sim$}}
    \raise1pt\hbox{$>$}}}         

\renewcommand{\d}{{\rm d}}
\newcommand{\eff}{{\rm eff}}

\makeatletter
\def\l@subsubsection#1#2{}
\makeatother

\newcommand{\sapienza}{Dipartimento di Fisica, Sapienza Università 
	di Roma, Piazzale Aldo Moro 5, 00185, Roma, Italy}
\newcommand{\infn}{INFN, Sezione di Roma, Piazzale Aldo Moro 2, 00185, Roma, Italy}

\begin{document}
\title{
Tidal Love numbers and approximate universal relations for fermion soliton stars
}

\begin{abstract}
Fermion soliton stars are a consistent model of exotic compact objects which involve a nonlinear interaction between a real scalar field and fermions through a Yukawa term. This interaction results in an effective fermion mass that depends upon the vacuum structure in the scalar potential. In this work we investigate the tidal deformations of fermion soliton stars and compute the corresponding tidal Love numbers for different model parameters. Furthermore, we discuss the existence of approximate universal relations for the electric and magnetic tidal deformabilities of these stars, and compare them with other solutions of general relativity, such as neutron stars or boson stars. These relations for fermion soliton stars are less universal than for neutron stars, but they are sufficiently different from the ordinary neutron star case that a measurement of the electric and magnetic tidal Love numbers (as potentially achievable by next-generation gravitational wave detectors) can be used to disentangle these families of compact objects. Finally, we discuss the conditions for tidal disruption of fermion soliton stars in a binary system and estimate the detectability of the electromagnetic signal associated with such tidal disruption events.
\end{abstract}

\author{Emanuele Berti}
\email{berti@jhu.edu}
\affiliation{Department of Physics and Astronomy, Johns Hopkins University, 3400 North Charles Street, Baltimore, Maryland 21218, USA}

\author{Valerio De Luca}
\email{vdeluca@sas.upenn.edu}
\affiliation{Center for Particle Cosmology, Department of Physics and Astronomy,
University of Pennsylvania 209 S. 33rd St., Philadelphia, PA 19104, USA}

\author{Loris Del Grosso}
\email{loris.delgrosso@uniroma1.it}
\affiliation{\sapienza}
\affiliation{\infn}

\author{Paolo Pani}
\email{paolo.pani@uniroma1.it}
\affiliation{\sapienza}
\affiliation{\infn}

\date{\today}
\maketitle

{
  \hypersetup{linkcolor=black}
}

\section{Introduction}
\label{sec:intro}

Astrophysical binary systems undergo tidal interactions that hold invaluable clues about the internal structure of compact objects. These interactions intricately shape the dynamics of binary sources, leaving distinct signatures in the signals they emit, detectable across both the gravitational-wave~(GW) and electromagnetic spectra~\cite{poisson_will_2014}. A robust analytical framework to understand tidal effects is encapsulated in the concept of tidal Love numbers~(TLNs), which quantify the deformability properties of self-gravitating bodies~\cite{1909MNRAS..69..476L}. Initially devised in the context of Newtonian gravity, TLNs have since been generalized in a fully relativistic context~\cite{Hinderer:2007mb,Binnington:2009bb,Damour:2009vw}.

Their significance has been particularly highlighted in the study of binary neutron star~(NS) mergers, offering tantalizing prospects for constraining the equation of state~(EOS) of dense matter through GW observations~\cite{Baiotti:2010xh,Baiotti:2011am,Vines:2011ud,Pannarale:2011pk,Vines:2010ca,Lackey:2011vz,Lackey:2013axa,Flanagan:2007ix,Favata:2013rwa,Yagi:2013baa,Maselli:2013mva,Maselli:2013rza,DelPozzo:2013ala,TheLIGOScientific:2017qsa,Bauswein:2017vtn, Most:2018hfd,Harry:2018hke,Annala:2017llu, Abbott:2018exr,Akcay:2018yyh,Abdelsalhin:2018reg, Jimenez-Forteza:2018buh, Banihashemi:2018xfb, Dietrich:2019kaq, Dietrich:2020eud, Henry:2020ski, Pacilio:2021jmq,Maselli:2020uol} (see~\cite{GuerraChaves:2019foa,Chatziioannou:2020pqz} for reviews). Remarkably, these objects exhibit nearly EOS-independent relations between their moment of inertia, spin-induced quadrupole moment, and electric quadrupolar tidal deformability, which are found to hold with about 1$\%$ accuracy~\cite{Yagi:2013bca,Yagi:2013awa,Yagi:2016ejg}. A similar approximate universality exists between TLNs of different multipolar order and different parity~\cite{Yagi:2013sva,Delsate:2015wia}.

The TLNs have also been studied in the context of asymptotically flat black holes~(BHs), with the intriguing finding that, in general relativity, they vanish in the limit of static external perturbations for BHs in isolation~\cite{Binnington:2009bb,Damour:2009vw,Damour:2009va,Pani:2015hfa,Pani:2015nua,Gurlebeck:2015xpa,Porto:2016zng,LeTiec:2020spy, Chia:2020yla,LeTiec:2020bos,Charalambous:2021mea,Charalambous:2021kcz,Ivanov:2022hlo,Charalambous:2022rre,Katagiri:2022vyz, Ivanov:2022qqt,Berens:2022ebl, Bhatt:2023zsy, Sharma:2024hlz}. 
However, this property is delicate, being violated for BH mimickers~\cite{Pani:2015tga,Cardoso:2017cfl}, in the presence of a cosmological constant~\cite{Nair:2024mya} or extended gravitational theories~\cite{Cardoso:2017cfl,Cardoso:2018ptl,DeLuca:2022tkm}, in higher dimensions~\cite{Kol:2011vg,Cardoso:2019vof, Hui:2020xxx,Rodriguez:2023xjd,Charalambous:2023jgq,Charalambous:2024tdj} or in 
nonvacuum environments in the presence of secular effects, such as accretion or  
superradiant instabilities of ultralight bosonic fields~\cite{Baumann:2018vus,Cardoso:2019upw,DeLuca:2021ite,Cardoso:2021wlq, DeLuca:2022xlz}. Furthermore, recently there has been an emerging interest in computing dynamical Love numbers~\cite{Nair:2022xfm, Saketh:2023bul, Perry:2023wmm, Chakraborty:2023zed} and incorporating nonlinear effects~\cite{DeLuca:2023mio, Riva:2023rcm}.

The forthcoming next-generation ground-based GW detectors~\cite{Kalogera:2021bya}, such as the Einstein Telescope~\cite{Maggiore:2019uih,Branchesi:2023mws} and Cosmic Explorer~\cite{Evans:2016mbw,Essick:2017wyl}, will improve the accuracy of measurements of the tidal deformability~\cite{JimenezForteza:2018rwr,Pacilio:2021jmq}, potentially unveiling the existence of new physics in the gravitational signals. 
This possibility comprises alternative end-states of gravitational collapse known as ``exotic compact objects''~(ECOs). Some of the simplest ECO models include self-gravitating solitons, such as boson stars, which are stable solutions of the Einstein-Klein-Gordon theory with a complex and massive scalar field~\cite{Jetzer:1991jr,Schunck:2003kk,Liebling:2012fv} (as opposed to real scalar fields which, constrained by no-go theorems~\cite{Derrick:1964ww, Herdeiro:2019oqp}, give rise to time-dependent solutions known as oscillatons~\cite{Seidel:1991zh}). Similar solitonic configurations with nonzero spin fields have been found. These include Dirac stars~\cite{Finster:1998ws}, which stem from solutions of the Einstein-Dirac equations with two neutral fermions, or Proca stars~\cite{Brito:2015pxa}, self-gravitating configurations supported by a complex spin-1 field. The TLNs of these objects have been previously investigated in Refs.~\cite{Cardoso:2017cfl,Herdeiro:2020kba,Chen:2023vet}. 

In this work, we focus on a class of solitonic solutions known as fermion soliton stars (FSS). These are solutions of general relativity featuring a real scalar field, with two (non)degenerate vacua, coupled to massive fermions via a Yukawa term~\cite{Lee:1986tr, DelGrosso:2023trq, DelGrosso:2023dmv}. For sufficiently strong coupling, the fermions deform the true vacuum state and create energetically preferred false-vacuum pockets wherein fermions are trapped~\cite{DelGrosso:2024wmy}. This scenario may play a role in various contexts in and beyond the Standard Model, providing a support mechanism for new compact objects that can form in the early Universe and serve as dark matter candidates. Remarkably, for natural model parameters between the QCD and the electroweak scale, this model predicts the existence of compact objects in the subsolar/solar range~\cite{DelGrosso:2024wmy}, which could be relevant for current and future LIGO-Virgo-KAGRA observations (see e.g. Refs.~\cite{Crescimbeni:2024cwh,Golomb:2024mmt}).

Here we study the tidal deformability and the quadrupolar TLNs, both of electric-type and of magnetic-type, of spherically symmetric FSSs. We derive the perturbation equations and perform numerical calculations of the Love numbers corresponding to specific background solutions. We observe that both classes of Love numbers are qualitatively similar to those of NSs. Finally, we investigate the existence of universality relations between the TLNs in the two sectors.

The manuscript is organized as follows.
In Sec.~\ref{sec:background} we outline the model and review the FSS solutions.
In Sec.~\ref{sec:perturbations} we give the equations governing the perturbations.
In Sec.~\ref{sec:results} we present our main findings.
In Sec.~\ref{sec:conclusions} we discuss the implications of our results and possible directions for future work.
In the following we use the metric signature $(-,+,+,+)$ and natural units ($\hbar = c = 1$). 

\section{Background}
\label{sec:background}

We are interested in FSSs~\cite{Lee:1986tr, DelGrosso:2023dmv, DelGrosso:2023trq}, self-gravitating solutions of general relativity in the presence of a real scalar field coupled to a fermion field via a Yukawa term. The action of the theory reads
\begin{align}\label{theory_fund}
    S = \int \d^4 x \sqrt{-g}
    \Big[
    &\frac{R}{16\pi G} - \unmezzo \partial^\mu \phi \partial_\mu \phi - U(\phi)
    \nonumber \\
    &-\bar{\psi}\gamma^\mu D_\mu\psi - (m_f - f\phi )\bar{\psi} \psi\Big],
\end{align}
where $R$ is the Ricci scalar of the metric $g_{\mu\nu}$, $\phi$ is the scalar field with potential $U(\phi)$, $\psi$ is the fermion with mass $m_f$, and $f$ is the Yukawa coupling.
With the normalization used for the fermionic kinetic term, the Dirac matrices have an extra $-i$ factor with respect to the usual definition, but satisfy the usual relation $\{\gamma^\mu, \gamma^\mu\} = 2 g^{\mu\nu}$.
The Yukawa coupling provides an effective mass, $m_{\rm eff}=m_f-f\phi$, that is crucial for the existence of these solutions~\cite{Lee:1986tr,DelGrosso:2023trq}, which circumvent classical no-go theorems for the existence of solitons~\cite{Derrick:1964ww,Herdeiro:2019oqp}.
The covariant derivative $D_\mu$ in Eq.~\eqref{theory_fund} takes into account the spin connection of the fermionic field. 

The scalar quartic potential reads
\begin{align}\label{potential_fund}
  U(\phi) = \frac{\mu^2 v_\F^2}{12} \frac{v_\F}{v_\B}\Big( \frac{\phi}{v_\F} \Big)^2 \Big[& 3 \Big( \frac{\phi}{v_\F} \Big)^2 \nonumber\\   - 4 \Big( \frac{\phi}{v_\F} \Big) 
  & \Big(1 + \frac{v_\B}{v_\F}\Big) + \frac{6 v_\B}{v_\F}\Big] \, ,
\end{align}
and features two minima at $\phi=0$ and $\phi=v_\F$, separated by a maximum located at $\phi = v_\B$. The parameter $\mu$ is the mass of the scalar field. By defining $\zeta = v_\B / v_\F$, it is possible to control the energy difference between vacua. When $\zeta = 1/2$ the two minima are degenerate. If $\zeta > 1/2$, the minimum at $\phi = v_\F$ has more energy than the minimum at $\phi = 0$; the opposite happens for $\zeta < 1/2$.

We will focus on scenarios in which the fermion becomes effectively massless (i.e., $m_\eff = 0$) when the scalar field corresponds to the second vacuum, $\phi=v_\F$. This condition implies fixing 
\begin{equation}\label{eq:f_fixing}
	f  =\frac{m_f}{v_\F}.
\end{equation}
At the background level, we consider spherically symmetric equilibrium configurations described by the line element
\begin{equation}\label{eq:general_spacetime}
\d s^2 = 
-e^{2u(\rho)} \d t^2 
+ e^{2v(\rho)}\d \rho^2 
+ \rho^2 (\d \theta^2 + \sin^2\theta \d \varphi^2),
\end{equation}
in terms of two real metric functions $u(\rho)$ and $v(\rho)$.
We will assume that the background scalar field is also static and spherically symmetric, $\phi(t,\rho,\theta,\varphi) = \phi(\rho)$.

\begin{figure}[t] 
\centering
\includegraphics[width=0.78\linewidth]{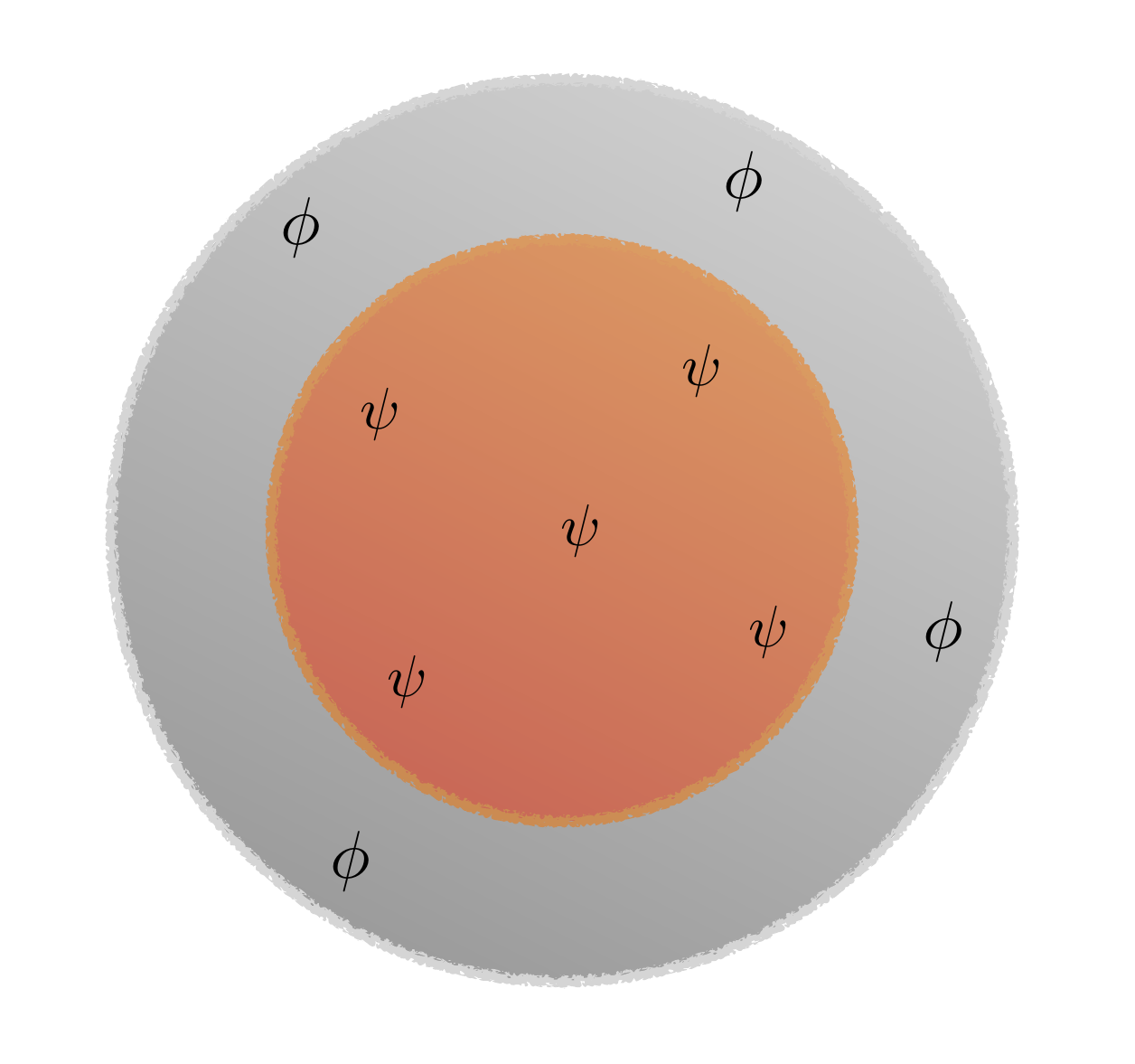}%
\caption{
   Pictorial illustration of a FSS.
   The inner region is dominated by fermions, smoothly connected, and surrounded by, an outer layer made of the scalar field.
   }
   \label{fig:illustration}
\end{figure}

\begin{figure*}[t!]
	\centering
 \includegraphics[width=0.45\textwidth]{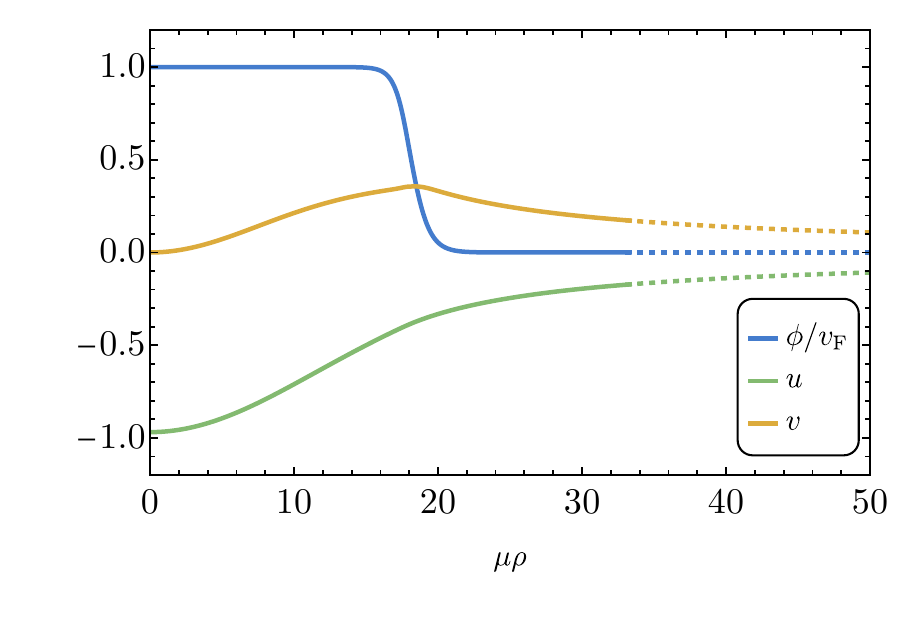}
 \includegraphics[width=0.45\textwidth]{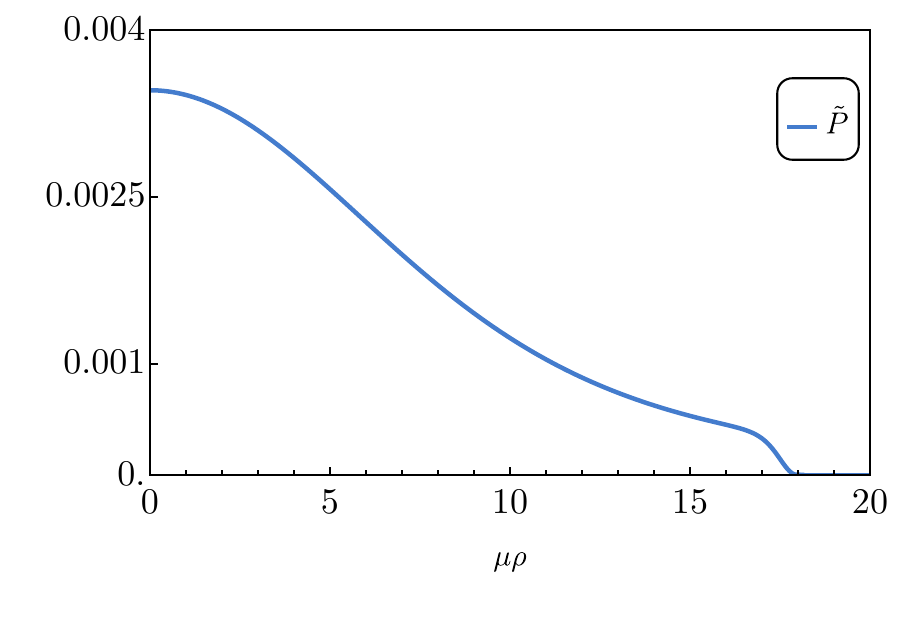}
  \includegraphics[width=0.45\textwidth]{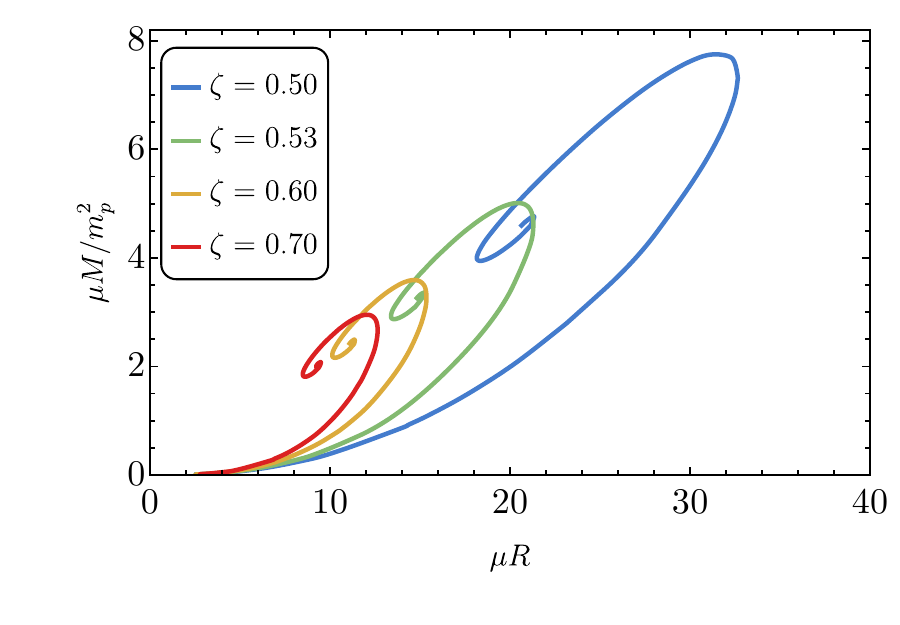}
   \includegraphics[width=0.45\textwidth]{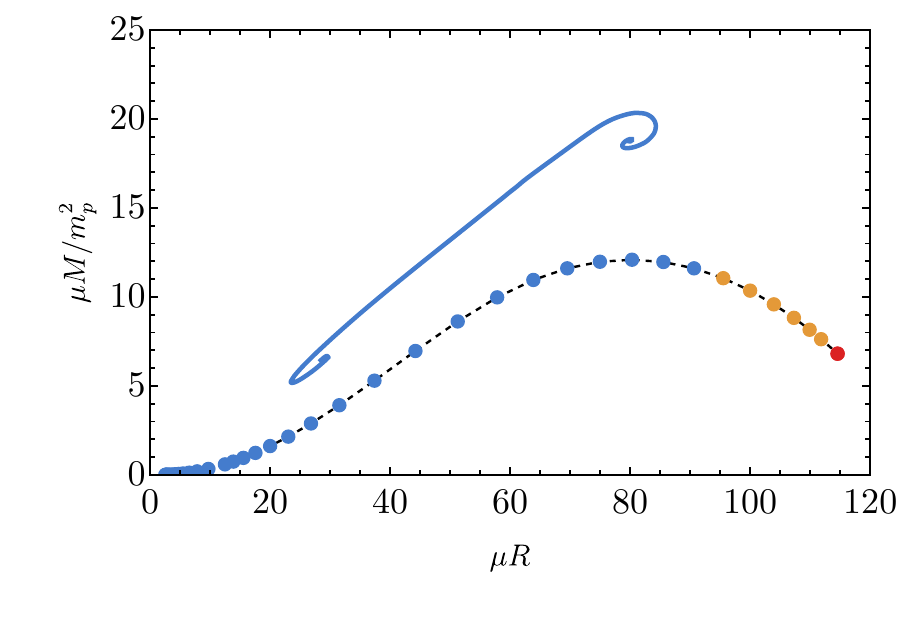}
	\caption{Top panels: Radial profiles of scalar field $\phi$, metric functions $u, v$ (left panel) and fermion
pressure (right panel) for a typical background configuration ($\Lambda = 0.15, \eta = 3, \zeta = 0.53$). Continuous lines represent numerical data,
whereas dashed lines reconstruct the asymptotic behavior of the solution by fitting with the Schwarzschild spacetime.
The mass and radius of this configuration are $\mu M / m_p^2 \approx 4.85$ and $\mu R \approx 19.04$, and the compactness is $C \approx 0.25$.
 Bottom panels: Mass-radius diagrams describing background solutions for FSSs with a positive effective cosmological constant in the interior (left panel, $\zeta \geq 0.5$) and a negative one (right panel, $\zeta = 0.49$). We fixed $\Lambda = 0.15$ and $\eta = 3$ as representative values. The blue curves/points in the right panel correspond to background solutions satisfying the weak energy condition, while the others violate it. The red circle corresponds to $\tilde{P}_c \to 0$, i.e., a purely-scalar solitonic configuration in the absence of fermions that does not exist in the $\zeta \geq 1/2$ case~\cite{DelGrosso:2023dmv}.}
\label{background_panel}
\end{figure*}

Fermions are treated in the Thomas-Fermi approximation~\cite{Lee:1986tr, DelGrosso:2023trq}, i.e., they enter Einstein's equations as a perfect fluid characterized by an energy-momentum tensor of the form
\begin{equation}
	T^{[f]}_{\mu\nu} = (W+P)u_\mu u_\nu + Pg_{\mu\nu},
\end{equation}
where $W$ is the energy density and $P$ is the pressure of the fluid, while they also enter the scalar field equation through the scalar density $S$. These quantities are defined as follows:
\begin{align}
\label{fermion_energy}
W &= \frac{2}{(2\pi)^3} \int_0^{k_{\rm F}} \d^3 k \, \epsilon_k,
\\
\label{fermion_pressure}
 P &= \frac{2}{(2\pi)^3} \int_0^{k_{\rm F}} \d^3 k \hspace{0.1cm} \frac{k^2}{3\epsilon_k},
\\
\label{fermion_density}
S &=  \frac{2}{(2\pi)^3} \int_0^{k_{\rm F}} \d^3 k \hspace{0.1cm} \frac{m_\eff}{\epsilon_k},
\end{align}
where $\epsilon_k = \sqrt{k^2 + m_\eff^2}$.
Notice that $W=W(x^\mu)$ through the spacetime dependence of $k_{\rm F}$ and $m_\eff$ (the same holds for $P$ and $S$). The integrals in Eqs.~\eqref{fermion_energy},~\eqref{fermion_pressure},~\eqref{fermion_density} can be computed analytically, as shown for example in Ref.~\cite{DelGrosso:2023trq}.

The fermion fluid is fully characterized once the Fermi momentum $k_\F$ is given.  At the background level, since the spacetime is static and spherically symmetric, $k_{\rm F} = k_{\rm F}(\rho)$ can only be a function of the radial coordinate. Minimizing the energy at a fixed number of fermions ensures~\cite{Lee:1986tr, DelGrosso:2023trq}
\begin{equation}
\label{Eqkf}
    k_{\rm F}^2(\rho) = \omega_{\rm F}^2e^{-2u(\rho)} - (m_f - f\phi(\rho))^2\,,
\end{equation}
where $\omega_{\rm F}$ is the Fermi energy at the origin ($\rho=0$), which can be written in terms of the fermion central pressure $P(\rho = 0) \equiv P_c$ (see Ref.~\cite{DelGrosso:2023trq} for details).

In order to simplify the numerical integrations, and to develop some physical intuition, it is convenient to write the field equations in terms of dimensionless quantities.
To this end, we define 
\begin{equation}\label{dimensionless_variables}
x  = \frac{k_{\rm F}}{m_f}, \qquad
y  = \frac{\phi}{v_\F}, \qquad
r = \rho \mu .  
\end{equation}
Therefore, the potential $U$ and kinetic term $V=  \unmezzo e^{-2v(\rho)} (\partial_\rho \phi)^2$ become
\begin{align}\label{dimensionless_scalar_quantities}
U  
\equiv  \mu^2 v_\F^2 \,\tilde{U}(y) ,
\qquad 
V 
\equiv \mu^2 v_\F^2 \,\tilde{V}(y)\,.
\end{align}
Moreover, we introduce the following dimensionless fermionic quantities:
\begin{equation}\label{dimensionless_fermionic_quantities}
\tilde{W} = \frac{W}{m_f^4}, \qquad
\tilde{P} = \frac{P}{m_f^4}, \qquad
\tilde{S} = \frac{S}{m_f^3} .  
\end{equation}
It is convenient to further introduce the dimensionless combination of parameters
\begin{align}\label{dimensionless_parameters}
	\Lambda &= \frac{\sqrt{8\pi}v_\F}{m_p}, 
	\qquad 
	\eta  = \frac{m_f}{\mu^{1/2} v_\F^{1/2}},
\end{align}
where $m_p=1/\sqrt{G}$ is the Planck mass.

Finally, the field equations (i.e.\ the Einstein-Klein-Gordon equations with the addition of the Fermi momentum equation) take the compact form
\begin{align}\label{fund_sistema_dimensionlesse_dimensionlesse}
& e^{-2v}-1-2  e^{-2v} r\partial_r v = -\Lambda^2 r^2  \left [ \eta^4 \tilde{W} + \tilde{U} +  \tilde{V} \right],
\nn 
\\
&
e^{-2 v} - 1 + 2  e^{- 2v} r\partial_r u =\Lambda^2 r^2  \left [\eta^4 \tilde{P} - \tilde{U} +  \tilde{V}\ \right],
\nn 
\\
&
e^{-2v}\Big[  \partial_r ^2 y +  \Big(\partial_r u - \partial_r v + \frac{2}{r}\Big)\partial_r y \Big] 
		= \frac{\partial \tilde{U}}{\partial y} - \eta^4\tilde{S} ,
		\nn 
		\\
& 
x^2  
= \tilde{\omega}_{\rm F}^2 e^{-2 u (r)} - (1-y)^2,
\end{align}
where $\tilde U$, $\tilde V$, $\tilde P$, $\tilde W$, and $\tilde S$ depend on $x$, $y$, and $r$, and we also introduced 
$\tilde{\omega}_{\rm F} = {\omega_{\rm F}}/{m_f}$.
Static and spherically symmetric configurations are solutions of the above system of ordinary differential equations. More details about the boundary conditions and the numerical procedure to obtain these solutions can be found in Refs.~\cite{DelGrosso:2023trq, DelGrosso:2023dmv}.

A pictorial representation of these compact objects is shown in Fig.~\ref{fig:illustration}. The inner region, dominated by fermions, is surrounded by an outer layer from the scalar field component. As the scalar field moves away from its central value at $\rho \to 0$, the fermion effective mass grows, resulting into a quick drop of the pressure. The corresponding macroscopic size $R$ of the star is therefore found to be very close to where the scalar field starts moving away from the false vacuum. We will define this radius as the region containing $99\%$ of the total mass $M$~\cite{DelGrosso:2023dmv}. 

In the top panels of Fig.~\ref{background_panel} we show an example of a background solution by plotting the radial profiles for the metric, scalar field, and fermion
pressure. We explicitly see that the scalar field is in a solitonic configuration interpolating between the false and the true vacuum of the theory.
Finally, in the bottom panels of Fig.~\ref{background_panel} we show the mass-radius diagrams of fermion soliton stars for different values of $\zeta$. For $\zeta < 0.5$ we highlight the existence of two disconnected branches. This pathological behavior is a consequence of the violation of the weak energy condition, and it can result into an ill-defined mass (see Ref.~\cite{DelGrosso:2023dmv} for a detailed analysis).

\section{Perturbations}
\label{sec:perturbations}

The tidal deformabilities of compact objects can be computed using perturbation theory. This amounts to considering fluctuations in both the background metric and in the matter content of the theory, which in this case consists of the scalar and fermionic fields. 
The perturbed metric at first order reads
\begin{equation}
g_{\mu\nu}=\bar{g}_{\mu\nu}+h_{\mu\nu}\,,
\label{metricpertdef}
\end{equation}
where $\bar{g}_{\mu\nu}$ is the background spacetime metric and $h_{\mu\nu} \equiv \delta g_{\mu\nu}$ is a small tensorial perturbation. In the following we use a bar superscript to denote the unperturbed quantities. 

We assume that the perturbations are sourced by an external stationary tidal field. Consequently, all the perturbations of the metric/fluid are independent of time.

The spherical symmetry of the system allows us to decompose the first-order perturbation $h_{\mu\nu}$ in spherical harmonics and to separate the perturbation into even (polar) and odd (magnetic) parity sectors, $h_{\mu\nu}=h_{\mu\nu}^{\text{even}}+h_{\mu\nu}^{\text{odd}}$. In the Regge-Wheeler gauge, $h_{\mu\nu}$ is decomposed as~\cite{Regge:1957td}
\begin{widetext}
\begin{align}
\label{heven}
&h_{\mu\nu}^{\text{even}}=
\left(\begin{array}{cccc}
e^{2u}H_0^{lm}(\rho)Y^{lm} & H_1^{lm}(\rho)Y^{lm} & 0 & 0 \\
 H_1^{lm}(\rho)Y^{lm} & e^{2v}H_2^{lm}(\rho)Y^{lm} & 0 &0\\
 0 &0 & \rho^2 K^{lm}(\rho)Y^{lm}& 0\\ 
0&0 &0 & \rho^2\sin^2{\theta} K^{lm}(\rho)Y^{lm}
\end{array}\right)
,\\
\nonumber\\
\label{hodd}
&h_{\mu\nu}^{\text{odd}}=\left(\begin{array}{cccc}
0 & 0 & h_0^{lm}(\rho)S_\theta^{lm} & h_0^{lm}(\rho)S_\varphi^{lm} \\
0 &0 & h_1^{lm}(\rho)S_\theta^{lm} & h_1^{lm}(\rho)S_\varphi^{lm} \\
h_0^{lm}(\rho)S_\theta^{lm} &h_1^{lm}(\rho)S_\theta^{lm}  &0 &0\\
h_0^{lm}(\rho)S_\varphi^{lm}& h_1^{lm}(\rho)S_\varphi^{lm} & 0 &0 \end{array}
\right),
\end{align}
\end{widetext}
with scalar and odd vector harmonics $\left(S_\theta^{lm},S_\varphi^{lm}\right)\equiv \left(-Y_{,\varphi}^{lm}/\sin{\theta},\,\sin{\theta} \,Y_{,\theta}^{lm}\right),$ and assuming an implicit sum over the angular indices $l, m$. 
Similarly, we decompose the fluid perturbations in spherical harmonics as $\delta X = X_{1} (\rho) Y^{l m}$, where $\delta X$ indicates a given matter perturbation and $X_1 (\rho)$ its radial dependence (omitting the multipole dependence on $l,m$ and the sum over these indices). This results in an analogous decomposition for the fluid stress-energy tensor:
\begin{equation}
T_{\mu \nu} = \overline{T}_{\mu \nu} + \delta T_{\mu \nu}\,.
\end{equation}
The corresponding Einstein equations then read $\delta G_{\mu \nu} = 8 \pi G \delta T_{\mu \nu}$, in terms of the perturbed part of the Einstein tensor. Since the background spacetime is spherically symmetric, the two sectors are decoupled and can be solved independently. 

In the following we start by discussing the matter content, and then we compute the metric perturbations in the two sectors.

\subsection{Fermionic perturbations}\label{sec:fermionic_perturbations}
By assuming that also the perturbed fluid is perfect (i.e., by neglecting anisotropic stress), one can introduce the perturbed quantities $W = \Bar{W} + \delta W, P = \Bar{P} + \delta P, u^\mu = \Bar{u}^\mu + \delta u^\mu$, to get
\begin{equation}
\label{eq:perturbed_fermion_tensor}
    \delta T^{[f] \mu}_\nu = (\delta W + \delta P)\bar{u}^\mu \bar{u}_\nu + (\Bar{W}+ \Bar{P}) (\delta u^\mu \Bar{u}_\nu + \Bar{u}^\mu \delta u_\nu ) + \delta P \delta^\mu_\nu\,,
\end{equation}
where the background four-velocity of the fluid is simply $\Bar{u}^\mu = (e^{-u}, 0, 0, 0)$, while $\Bar{u}_\mu = (-e^u,0,0,0)$. It is important to stress that $\Bar{u}^\mu \propto \delta^\mu_0$, so the only nonzero, nondiagonal elements of Eq.~\eqref{eq:perturbed_fermion_tensor} are the $0i$ components. On the other hand, in the presence of anisotropic stress, the $ij$ contributions can also be different from zero.

Imposing $g_{\mu\nu} u^\mu u^\nu = -1$, one gets
\begin{equation}
\label{eq:fourvelocity_constraint}
    \delta g_{\mu\nu} \Bar{u}^\mu\Bar{u}^\nu + 2 \Bar{g}_{\mu\nu} \Bar{u}^\mu \delta u^\nu = 0 \Rightarrow \delta u^0 = \frac{1}{2} e^{-3u} \delta g_{00}\,,
\end{equation}
while $\delta u_0 =e^{-u} \delta g_{00}/2$. The $tt$-component of the fermionic perturbed energy-momentum tensor then reads
\begin{align}
    \delta T^{[f] 0}_0 &= -(\delta W + \delta P)+ (\Bar{W}+ \Bar{P}) [\delta u^0 (-e^u)  \nonumber \\
    &+ e^{-u} \delta u_0] + \delta P = -\delta W \,.
\end{align}
Similarly, it is straightforward to write down the remaining components of the perturbed stress-energy tensor:
\begin{align}
    \delta T^{[f] i}_j &= \delta P \delta^i_j \\
    \delta T^{[f] i}_0 &= -(\Bar{W}+\Bar{P})e^u \delta u^i  \\
    \delta T^{[f] 0}_i &= (\Bar{W}+\Bar{P})e^{-u}\delta u_i\,,
\end{align}
where $\delta u^i = d \xi^i / d\tau$, with $\xi^i$ the spatial displacement of the fluid element due to the perturbations, and $\tau$ the proper time. 

As discussed in Ref.~\cite{Pani:2018inf}, in the zero-frequency limit of the time-dependent response to an external gravitational perturbation the perturbed fluid can either be static (i.e., characterized by zero three-velocity)
or irrotational (i.e., characterized by zero vorticity).
The latter is a more realistic assumption in a binary system, as discussed in  Refs.~\cite{Landry:2015cva, Pani:2018inf, Delsate:2015wia}.
Such different fluid configurations are found to impact the magnetic TLNs in both their sign and value, while the electric TLNs remain the same in both cases~\cite{Pani:2018inf,Landry:2015cva}. Therefore, we write the spatial fluid velocity as~\cite{Abdelsalhin:2019ryu}
\begin{equation}
    \delta u^i = \{ 0, Q^{lm}(\rho) \, S_{\theta}^{lm}(\theta\,,\varphi), \frac{Q^{lm}(\rho)}{\sin{\theta}^2} \, S_{\varphi}^{lm}(\theta\,,\varphi)  \} \, ,
\end{equation}
where $i = \rho, \theta, \varphi$ and
\begin{equation}
    Q^{lm}(\rho) = - \frac{e^{-u}}{\rho^2} h^{lm}_0(\rho) \,,
\end{equation}
as required to describe an irrotational fluid~\cite{Kojima:1992ie}.

As already highlighted in the previous section, in the Thomas-Fermi approximation the fermionic fluid is fully characterized once the Fermi momentum $k_\F$ is given at each spacetime point. In the following we will assume a similar perturbative decomposition
\begin{equation}
    k_\F = \Bar{k}_\F + \delta k_\F\,.
\end{equation}
Adopting the same decomposition also for the scalar field $\phi = \Bar{\phi} + \delta \phi$, as discussed in the next subsection, we can write down $\delta W, \delta P, \delta S$ in terms of $\delta k_\F$ and $\delta \phi$, assuming that $W, P, S$ can be computed as in the background case: see Eqs.~\eqref{fermion_energy}, \eqref{fermion_pressure}, and \eqref{fermion_density}.
This assumption is valid as long as the time scale of the tidal perturbation is much longer than that of the fundamental fermionic degrees of freedom, which is always the case.

In order to close the system of equations, we need a further condition which relates $\delta k_\F$ to the metric and scalar functions, analogous to Eq.\eqref{Eqkf} for the background quantities. In principle, such a condition can be derived by generalizing the Thomas-Fermi approximation to a generic spacetime. This task can actually be accomplished by using the Einstein equations. In particular, adding up the $\theta \theta$ and $\varphi \varphi$ components gives $\delta P$ as a function of $H_0$ and $\delta \phi$. We can also use Eq.~\eqref{fermion_pressure} to write
\begin{equation}
\label{eq:pertubed_pression}
    \delta P = \frac{\delta P}{\delta k_\F} \delta k_\F +  \frac{\delta P}{\delta \phi} \delta \phi \,.
\end{equation}
At this point we have two independent expressions for $\delta P$, which in turn gives $\delta k_F$ as a function of metric, scalar, and background quantities only.
In analogy with the background case, we therefore need to solve only the Einstein equations and the scalar field equation. The additional perturbations $\delta W$ and $\delta S$ are obtained by expanding Eqs.~\eqref{fermion_energy} and~\eqref{fermion_density} as done in Eq.~\eqref{eq:pertubed_pression}. We give the explicit expressions in Appendix~\ref{appendixA}.

\subsection{Scalar perturbations}

The energy-momentum tensor of a real scalar field reads 
\begin{align}
T_{\mu \nu}^{[\phi]} & = \partial_\mu \phi \partial_\nu \phi - g_{\mu \nu} \lp \frac{1}{2} g^{\alpha \beta} \partial_\alpha \phi \partial_\beta \phi + U(\phi) \rp\,,
\end{align}
such that its perturbation becomes
\begin{align}
& \delta T_{\mu \nu}^{[\phi]}  = \partial_\mu \delta \phi \partial_\nu \bar{\phi} + \partial_\nu \delta \phi \partial_\mu \bar{\phi}
-  \delta g_{\mu \nu} \lp \frac{1}{2} \bar{g}^{\alpha \beta} \partial_\alpha \bar{\phi} \partial_\beta \bar{\phi} + U(\bar{\phi}) \rp \nonumber \\
& - \bar{g}_{\mu \nu} \lp \frac{1}{2} \delta g^{\alpha \beta} \partial_\alpha \bar{\phi} \partial_\beta \bar{\phi} + \bar{g}^{\alpha \sigma} \partial_\alpha \delta \phi \partial_\beta \bar{\phi} + \delta U\rp\,,
\end{align}
where we have decomposed the scalar potential as $U(\phi) = U(\Bar{\phi}) + \delta U$, and
\begin{equation}
     \delta U= \frac{\partial U(\phi)}{\partial \phi}\Big|_{\Bar{\phi}} \,\delta \phi \,.
\end{equation}

\subsection{Perturbation equations}

We now turn to the solution of the linearized Einstein equations, focusing first on the polar sector and then on the axial sector.

\subsubsection{Polar perturbations}
Taking the difference between the $\theta \theta$ and $\varphi \varphi$ component reveals that $H_2(\rho) = H_0(\rho)$. From $\delta G^t_\rho = 8\pi G \delta T^t_\rho$ we get $H_1 \equiv 0$. Also, the $\rho \theta$ component can be used to relate $K'$ to $\phi_1, H_0$, and $H_0'$ as follows:
\be
K' = H_0'+2 H_0 u'-16 \pi G \, \phi_1 \phi_0'\,.
\ee
Finally, the difference between the $\rho\rho$ and $tt$ components can be written as a master equation for $H_0$, with no further $K$ dependence:
\begin{widetext}
\centering
\begin{align}
\label{polar}
    H_0'' &+ 8\pi G\, e^{2v} (P_1 + W_1) + \Big( \frac{2}{\rho} + u' - v' \Big) H'_0 + 16\pi G \, \Big[  (u' + v' - \frac{2}{\rho})\Bar{\phi}' - \Bar{\phi}'' \Big] \phi_1 \\ & + \Big[\frac{2}{\rho^2} - \frac{e^{2v} (l^2 +l + 2)}{\rho^2} + 16 \pi G\, e^{2v} (\Bar{P} + \Bar{W} - \frac{1}{2} \Bar{\phi}'\hspace{0.01cm}^2 e^{-2v}) + \frac{4 u'}{\rho} - 4 u'\hspace{0.01cm}^2 \Big] H_0  = 0
     \nonumber  \, ,
\end{align}
\end{widetext}
where $P_1$, $W_1$ and $U_1$ are the radial components of the matter perturbations $\delta P$, $\delta W$ and $\delta U$ after decomposition in spherical harmonics.
The equation of motion for the scalar field reads
\begin{widetext}
\centering
\begin{align}
& \phi_1'' + \phi_1' \lp u'-v'+\frac{2}{\rho} \rp + H_0 \left[\lp u'+v'-\frac{2}{\rho}\rp \Bar{\phi}' -\Bar{\phi}''\right] \nonumber \\
&- \lp \frac{l(l+1) e^{2 v}}{\rho^2} + 16 \pi G \Bar{\phi}'\hspace{0.01cm}^2 \rp \phi_1 = e^{2 v}  \frac{\partial^2 U(\phi)}{\partial \phi^2}\Big|_{\Bar{\phi}}\, \phi_1  -  e^{2 v} f S_1\,,
\end{align} 
\end{widetext}
where, again, $\phi_1$ is the radial component of the scalar field perturbation $\delta \phi$ after expansion in spherical harmonics.

\subsubsection{Axial perturbations}

The $\rho \varphi$ component of the perturbed Einstein equations implies $h_1=0$, and the $\rho \theta$ component implies $\phi_1=0$. We are therefore left with a single radial equation for the perturbed function $h_0$ (the $t \varphi$ component):
\begin{widetext}
\centering
\begin{align}
h_0'' - (u'+v')\,h_0' 
+\frac{\big[2-l(l+1)\big] e^{2v}-2 + 2 \rho (u'+v')
}{\rho^2}\,h_0 = 0\,.
\label{eqh0}
\end{align}
\end{widetext}
We numerically solve all the perturbed equations by applying the transformation given by Eqs.~\eqref{dimensionless_variables},~\eqref{dimensionless_scalar_quantities},~\eqref{dimensionless_fermionic_quantities},~\eqref{dimensionless_parameters}, in analogy with what we did for the background equations in Sec.~\ref{sec:background}.

\section{Tidal Love numbers}
\label{sec:results}

Nonspinning, spherically symmetric, FSSs immersed in an external stationary tidal field will be deformed and develop a multipolar structure in response to the external field. This phenomenon may occur in coalescing binary systems, where each component tidally deforms its companion because of gravity. The assumption of a stationary field holds only in the early inspiral phase, when the orbital separation from the companion is very large and the orbit is slowly varying in time. Furthermore, one can make use of the approximation that the multipolar deformation induced on the objects is linear in the strength of the external tidal field to define the TLNs as the ratio between the induced multipole moments and the tidal moments of the external gravitational field: 
\begin{align}
Q_L = \lambda_l G_L, \qquad S_L = \sigma_l H_L, \qquad l \gtrsim 2\,. 
\end{align}
Here the symbols $Q_L$ ($S_L$) denote the mass (current) multipole moments of order $l$ of the object ($L$ being a multi-index containing a number $l$ of individual indices), and $G_L$ ($H_L$) the corresponding electric (magnetic) tidal multipole moments.

The parameters $\lambda_l$ ($\sigma_l$) are the electric (magnetic) tidal deformabilities, related to the dimensionless TLNs $k^E_l$ ($k^M_l$) through the relations~\cite{Hinderer:2007mb,Yagi:2013sva, Abdelsalhin:2019ryu}
\begin{equation}
\label{eq:lovetidal}
\begin{aligned}
\lambda_l & = (G M)^{2l+1} \bar{\lambda}_l = \frac{2 R^{2l+1}}{(2l-1)!!}     k^E_l\,, \\
\sigma_l & = (G M)^{2l+1} \bar{\sigma}_l = \frac{(l-1)R^{2l+1}}{4(l+2)(2l-1)!!}   k^M_l\,,
\end{aligned} \qquad l \geq 2 \,,
\end{equation}
in terms of the FSS mass $M$ and radius $R$. The main contribution to the star's deformation comes from the quadrupole ($l=2$), which will be the main focus of our analysis.

The TLNs can be extracted by asymptotically expanding the metric of the object, perturbed by the external tidal source, at spatial infinity.
In asymptotically Cartesian mass-centered coordinates, the time-time and time-space components of the metric read
\begin{widetext}
\centering
\begin{equation}
\begin{aligned}
\label{eq:metrictidal}
g_{00} = & -1 + \frac{2G M}{\rho} + \sum_{l \geq 2}  \left[  \frac{1}{\rho^{l+1}} \left( \frac{2(2l-1)!!}{l!} Q_L n_L \right)  + \rho^l \left(\frac{2}{l!} G_L n_L  \right) \right] \,,\\
g_{0i} = & \sum_{l \geq 2}  \left[  \frac{1}{\rho^{l+1}} \left( -\frac{4l(2l-1)!!}{(l+1)!} \epsilon_{ija_l} S_{jL-1} n_L   \right)  + \rho^{l} \left( \frac{l}{(l+1)!}\epsilon_{ija_l} H_{jL-1} n_L  \right) \right] \,,
\end{aligned}
\end{equation}
\end{widetext}
where $n_i = x_i/\rho$ is the unit radial vector, $n_L = n^{a_1}\dots n^{a_l}$, and we absorb any factor of $G$ in the definition of $Q_L$,~$S_L$~\cite{Thorne:1980ru, Abdelsalhin:2019ryu}. For simplicity we have neglected terms independent of $\rho$ and proportional to spherical harmonics of order $l' < l$. In this coordinate frame, the mass dipole of the object vanishes identically. From this expansion it is clear that the computation of the TLNs is based on the separation between the radially decaying multipolar response of the central object and the external growing solution. 

In the following subsections we compute the tidal deformabilities of nonrotating FSSs in both the polar (electric) and axial (magnetic) sectors. Since the background spacetime is spherically symmetric, these sectors are completely decoupled from each other and can be treated independently. Notice also that, while the electric response is the relativistic generalization of the Newtonian Love number, the magnetic sector is instead fully relativistic~\cite{Binnington:2009bb}, since current distributions do not gravitate in Newtonian theory.
Furthermore, one can expand the multipolar quantities in spherical harmonics, reducing the problem to radial equations, which are independent of the index $m$ and do not couple perturbations with different values of $l$. These equations must be solved in both the exterior and interior regions of the object, matching the solutions at a characteristic extraction radius $R_{\rm ext}$, as discussed below.

\begin{figure*}[t!]
	\centering
 \includegraphics[width=0.45\textwidth]{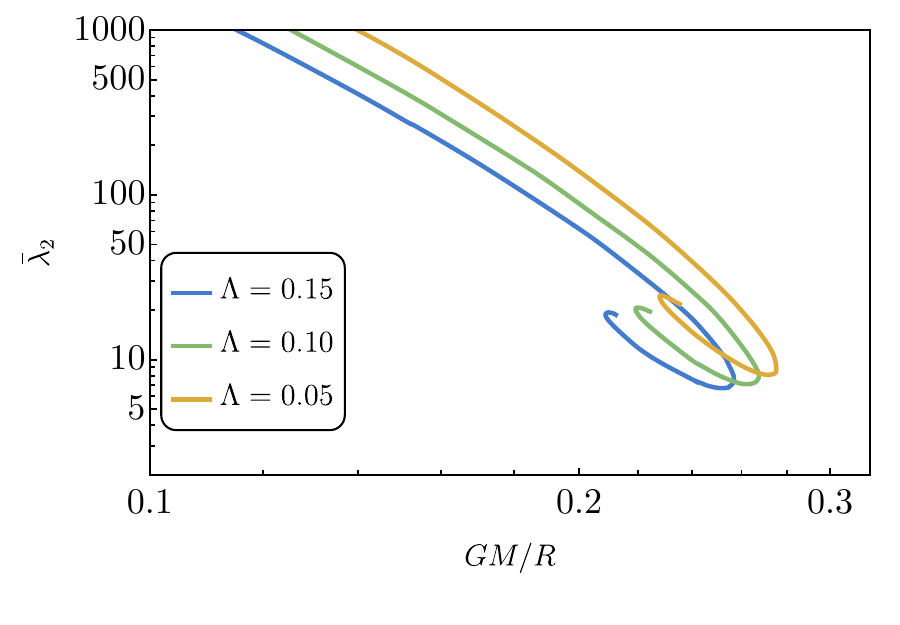}
 \includegraphics[width=0.45\textwidth]{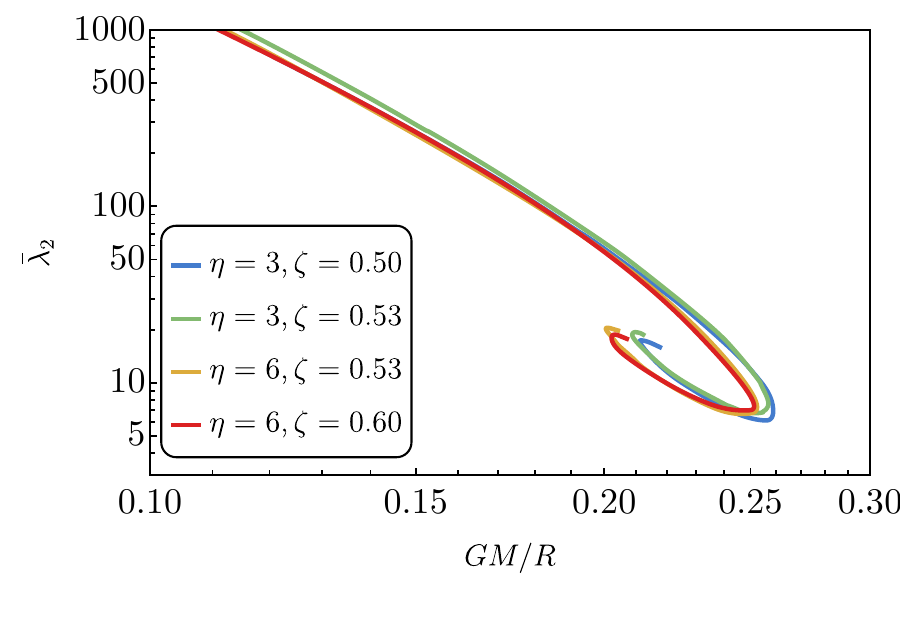}
  \includegraphics[width=0.45\textwidth]{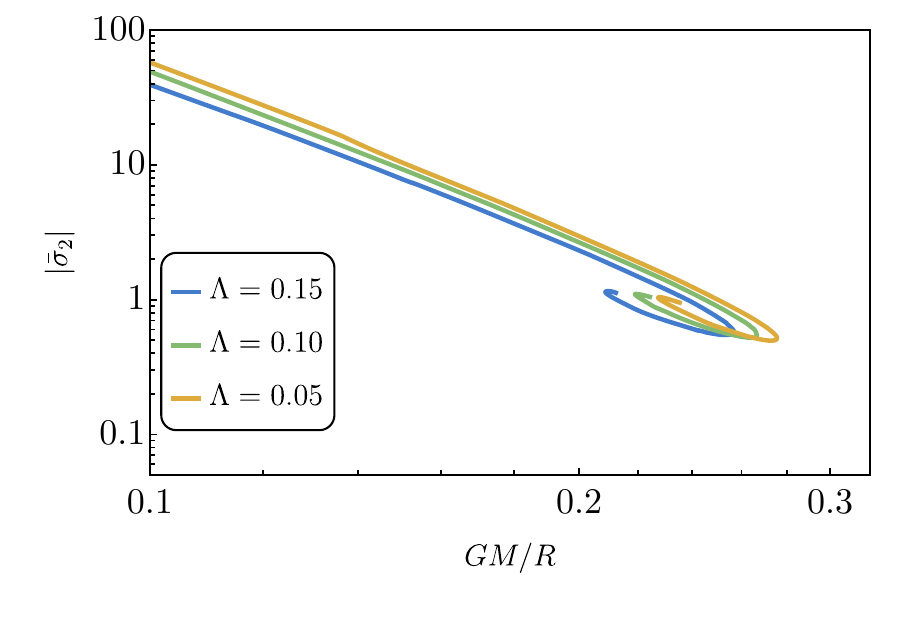}
   \includegraphics[width=0.45\textwidth]{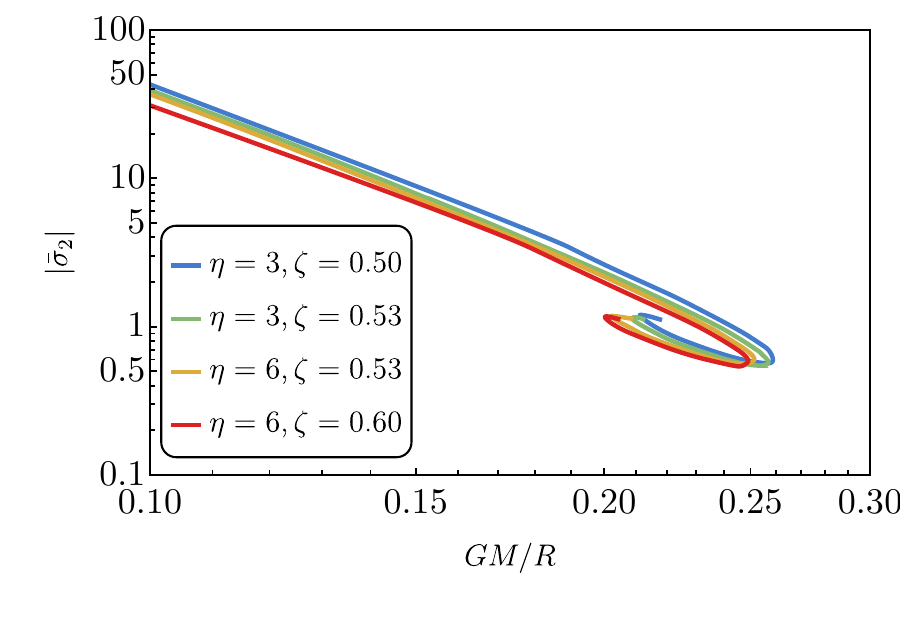}
	\caption{Electric (upper panels) and magnetic (lower panels) TLNs for the quadrupolar ($l =2$) mode as a function of the FSS compactness $G M / R$. The different curves correspond to different values of the model parameters. In the left panels, we fix $\eta = 3, \zeta = 0.53$ and vary $\Lambda$. In the right panels, we fix $\Lambda = 0.15$ and vary $\eta$ and $\zeta$. We explicitly see that the curves depend only mildly on $\eta$ and $\zeta$.}
	\label{FSSTLN}
\end{figure*}

\begin{figure*}[t!]
	\centering
 \includegraphics[width=0.45\textwidth]{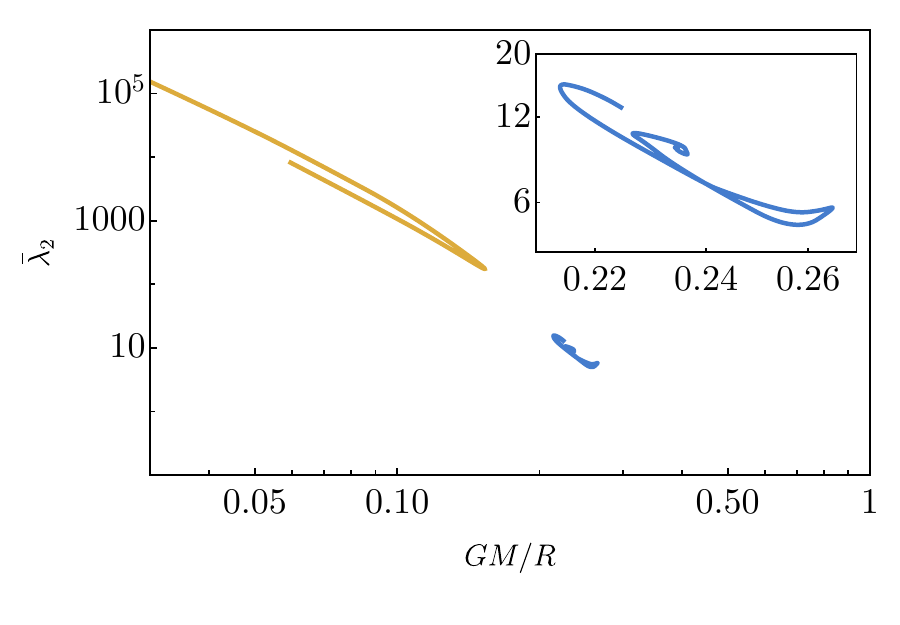}
  \includegraphics[width=0.45\textwidth]{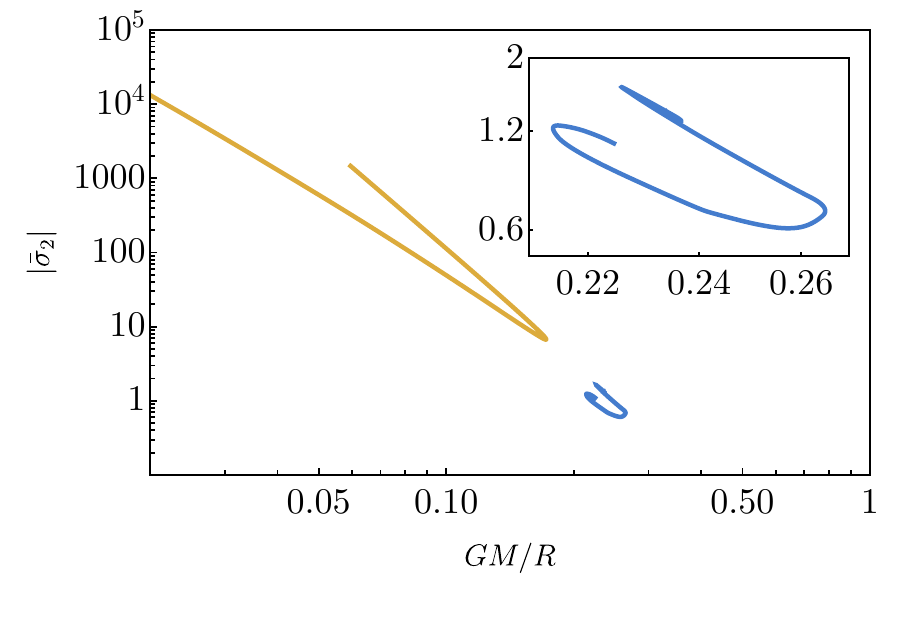}
	\caption{Same as Fig.~\ref{FSSTLN}, but for a negative effective cosmological constant ($\zeta = 0.49$) and  for $\Lambda = 0.15, \eta = 3$. In this case there are two disjoint branches of solutions, related to the corresponding branches in the mass-radius diagram shown in Fig.~\ref{background_panel} (the blue/orange lines corresponding to the upper/lower branches, respectively)~\cite{DelGrosso:2023trq}. The insets are zoomed-in versions of the small islands at large compactness, shown in blue.} 
	\label{FSSTLNzeta49}
\end{figure*}

\subsection{Polar sector}
In the polar sector, the perturbation equation for the field $H_0$ is given in Eq.~\eqref{polar}. In the vacuum region outside the object, it reduces to
\begin{align}
& H_0'' + \frac{2 (\rho-G M) }{\rho (\rho-2 G M)}H_0' \nonumber \\
& - \frac{  \left(4 G^2 M^2- 2 l (l+1) G M \rho+ l(l+1) \rho^2\right)}{\rho^2 (\rho-2 G M)^2}H_0 =0\,.
\end{align}
The vacuum solution is given in terms of associated Legendre polynomials as
\begin{equation}
\label{eq:extpolar}
H_0(\rho) = c_P \,  P_{l2}\left( \frac{\rho}{GM}-1 \right) + c_Q \, Q_{l2}\left( \frac{\rho}{GM}-1 \right)\,,
\end{equation}
where the integration constants $c_P$ and $c_Q$ are found in terms of $H_0(R_{\rm ext})$ and $H'_0(R_{\rm ext})$ by matching to the interior solution.
At spatial infinity, $\rho \to \infty$, one gets
\begin{equation}
\label{eq:h0ext}
H_0(\rho) \simeq  \tilde{c}_P \,\rho^{l} + \tilde{c}_Q \, \frac{1}{\rho^{l+1}} + \mathcal{O}\left(\frac{GM}{\rho}\right)\,,
\end{equation}
where the tilde is used to distinguish these coefficients from the ones introduced in Eq.~\eqref{eq:extpolar}, since an additional dependence on the mass $M$ and numerical factors arise from the asymptotic expansion. One can then plug this expansion into the $g_{00}$ component of the metric to get
\begin{equation}
\label{eq:extlarger}
g_{00} \sim -1 + \frac{2GM}{\rho} + \sum_{l \geq 2,m} \left( \frac{1}{\rho^{l+1}} \, \tilde{c}_{Q,lm} + \rho^{l} \,  \tilde{c}_{P,lm}     \right) Y^{lm}\,,
\end{equation}
which can be compared with the asymptotic expansion shown in Eqs.~\eqref{eq:metrictidal} once the multipole moments are properly decomposed in terms of symmetric trace-free (STF) tensors $Q_{lm}$ and $G_{lm}$.

By performing this matching, one can identify the growing solution in $H_0(\rho)$ with the tidal field and the decaying one with the response of the object, respectively, and then extract their multipole moments in terms of the coefficients $c_{Q,lm}$ and $c_{P,lm}$. For the leading multipole moment  $l =2$, the electric TLN reads~\cite{Hinderer:2007mb}
\small
\begin{widetext}
\centering
\begin{align}\label{eq:k2E}
k_2^E = \frac{8 (1-2 C)^2 C^5 (2 C (y-1)-y+2)}{10 C (C (2 C (C (2 C (y+1)+3 y-2)-11 y+13)+3 (5 y-8))-3 y+6)+15 (1-2 C)^2 (2 C (y-1)-y+2) \log (1-2 C)}\,,
\end{align}
\end{widetext}
\normalsize
where we have defined $C = G M/R_{\rm ext}$ and $y = \rho H_0'/H_0$, both evaluated at the extraction radius $R_{\rm ext}$, which is taken to be much larger than the FSS effective size $R$ (in order for the TLN to be independent from it). In the actual numerical computation, we could not go further than $R_{\rm ext}\sim 2R$ due to the high fine-tuning required by the shooting method used to compute the initial displacement of the scalar field. However, we checked that this yields sufficiently accurate numerical results, due to the exponential decay of the scalar field at $\rho>R$.

The matching variable $y$ can be computed by integrating Eq.~\eqref{polar} in the  interior of the FSS, imposing the boundary condition of regularity at the origin, $\rho=0$:
\begin{equation}
\label{eq:H0init}
H_0(\rho) = a_0 \, \rho^{l} \left[1+ \mathcal{O} \left(\rho^2 \right) \right]\,, \qquad \rho \to 0 \,,
\end{equation}
where the constant $a_0$ does not affect the TLN, since the problem is linear and this constant enters in both the strength of the tidal field and the size of the induced multipolar deformation, and therefore it cancels out when computing their ratio.

In the upper panels of Fig.~\ref{FSSTLN} we show the dimensionless TLN for the leading quadrupolar ($l =2$) mode as a function of the compactness, as we vary different parameters of the model. Similarly to NSs, the characteristic mass-radius diagram (shown in Fig.~2 of Ref.~\cite{DelGrosso:2023dmv}) displays a turning point at large compactness, which shows up also in the TLNs. We observe that in the phenomenologically interesting range around $G M/R \simeq 0.2$ corresponding to the critical solution, the TLN reaches values of the order of $\mathcal{O}(100)$, growing at small compactness due to the dependence $\bar{\lambda}_2 \propto C^{-5}$.
The strongest dependence is on the parameter $\Lambda$, describing the scalar field vacuum expectation value: in general, lower values of $\Lambda$ give rise to larger TLNs. The TLNs are almost independent of the degeneracy parameter $\zeta$, which is varied to span both degenerate and nondegenerate vacua, if one considers vacua with positive energy.

When $\zeta < 0.5$, the interior region of the vacuum bubble has negative energy, mimicking an effective anti de-Sitter space~\cite{DelGrosso:2023dmv}. This case, plotted in the left panel of Fig.~\ref{FSSTLNzeta49} for $\zeta = 0.49$, shows a different behavior of the TLN, reflecting what happens at the background level for the mass-radius diagram (see the bottom right panel of Fig.~\ref{background_panel}). Indeed, the presence of two disjoint branches at the background level gives rise to two distinct branches of the TLNs. In particular, the small islands at large compactness (i.e., the blue curves in Fig.~\ref{FSSTLNzeta49} magnified in the insets) represent the TLNs of the upper branch of the mass-radius diagram (continuous blue line in the bottom right panel of Fig.~\ref{background_panel}), whereas the orange curves are associated to the lower branch (dotted line in the bottom right panel of Fig.~\ref{background_panel}). Turning points in the mass-radius diagram give rise to turning points in the TLNs.

\subsection{Axial sector}
In the axial sector, the perturbation equation for the field $h_0$ is given in Eq.~\eqref{eqh0}. As discussed in the previous section, in the following we will work under the assumption of an irrotational fluid also in the perturbed configuration. Under this assumption, the perturbation equation at large distances reduces to
\be
h_0'' + \frac{ (4 G M- l(l+1) \rho)}{\rho^2 (\rho-2 GM)} h_0 = 0\,.
\ee
In the external region, the solution of the perturbation equation reads
\begin{equation}
\label{eq:extaxials}
\begin{aligned}
h_0(\rho) & = d_P \left(\frac{\rho}{2GM} \right)^{l+1} \,_2F_1 \left(-l+1,-l-2,-2l;\frac{2GM}{\rho} \right) \\
& +  d_Q   \left(\frac{2GM}{\rho} \right)^l \,_2F_1 \left(l-1,l+2,2l+2;\frac{2GM}{\rho} \right)\,,
\end{aligned}
\end{equation}
in terms of the hypergeometric function $_2F_1 \left(a,b,c; x \right)$. 
The constants $d_P$ and $d_Q$ can be found by the matching procedure in terms of $h_0(R_{\rm ext})$ and $h_0'(R_{\rm ext})$. In the large distance regime $\rho \to \infty$ the asymptotic solution is 
\begin{equation}
\label{eq:h0ext2}
h_0(\rho)=  \tilde{d}_P \,\rho^{l+1} + \tilde{d}_Q \, \frac{1}{\rho^{l}} + \mathcal{O} \left( \frac{GM}{\rho} \right)\,,
\end{equation}
where, as we discussed in the electric case, the tilded coefficients include additional dependence on the mass and further numerical factors. From this expression one can obtain the $g_{0\varphi}$ component of the metric as
\begin{equation}
\label{eq:extlarger2}
g_{0\varphi} \sim  \sum_{l \geq 2,m} \left( \frac{1}{\rho^{l}} \, \tilde{d}_{Q,lm} + \rho^{l+1} \,  \tilde{d}_{P,lm}  \right) S_{\varphi}^{lm}  \,.
\end{equation}
By STF decomposing the spatial-temporal part of the metric in Eqs.~\eqref{eq:metrictidal}, one can again identify the growing and decaying modes of the solutions, and extract the magnetic TLNs. For the leading $l = 2$ mode one gets
\begin{widetext}
\centering
\begin{align}\label{eq:k2B}
k_2^M = \frac{96 C^5 (2 C (y-2)-y+3)}{10 C (C (2 C (C y+C+y)+3 (y-1))-3 y+9)+15 (2 C (y-2)-y+3) \log (1-2 C)}\,,
\end{align}
\end{widetext}
\normalsize
where we have defined the quantity $y = \rho h_0'/h_0$, evaluated at the extraction radius $R_{\rm ext}$. As in the electric case, $y$ can be found by numerically integrating Eq.~\eqref{eqh0}, with boundary conditions at the center of the FSS given by
\begin{equation}
\label{eq:h0init}
h_0(\rho) = b_0 \, \rho^{l+1} \left[1+ \mathcal{O} \left(\rho^2 \right) \right]\,, \qquad \rho \to 0 \,,
\end{equation}
where the constant $b_0$ cancels out in the definition of the TLN.

Similarly to the even sector, the lower panels of Fig.~\ref{FSSTLN} and the right panel of Fig.~\ref{FSSTLNzeta49} show the magnetic TLN for the quadrupolar mode ($l = 2$) for $\zeta\geq1/2$ and $\zeta<1/2$, respectively. One can appreciate the same trends observed in the electric case. The magnetic TLN is found to be about an order of magnitude smaller than the electric one, as observed for NSs~\cite{Binnington:2009bb,Delsate:2015wia}. 

\begin{figure*}[t!]
	\centering
 	\includegraphics[width=0.493\textwidth]{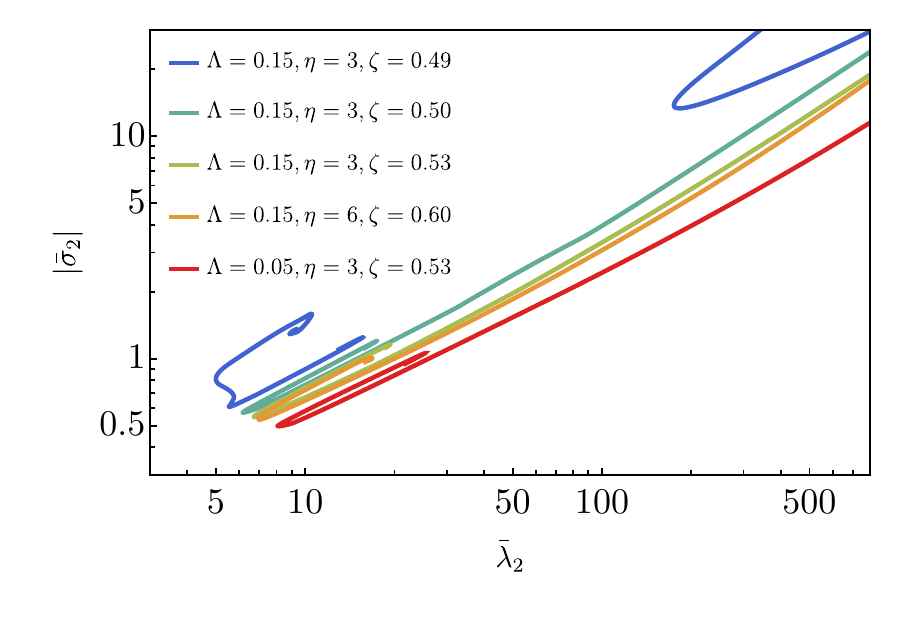}
	\includegraphics[width=0.493\textwidth]{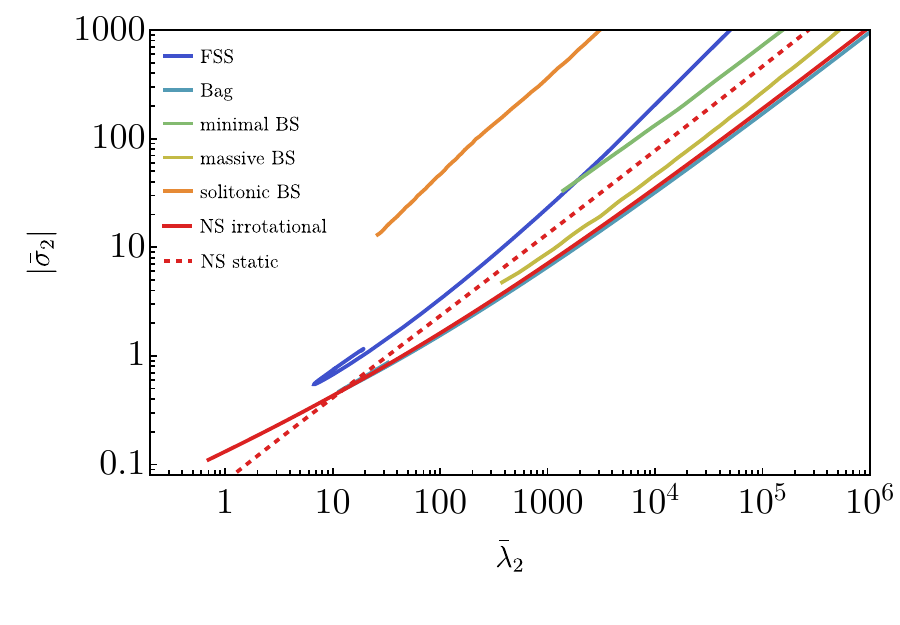}
	\caption{Left panel: universality relations for FSSs for different model parameters. In the $\zeta = 0.49$ case, the lower (upper) branch corresponds to the large compactness, blue (small compactness, orange) line of Fig.~\ref{FSSTLNzeta49}. Right panel: comparison of the FSS curve with $\Lambda = 0.15, \eta = 3, \zeta = 0.53$ with the universality curves for other compact objects, such as NSs in the irrotational (solid red line) and static configurations (dashed red line), soliton boson stars (orange line), massive boson stars (yellow line), minimal boson stars (green line) and an effective bag model (Bag, water green line).}
	\label{universalityNS}
\end{figure*}

\subsection{Quasi-universal relations}

The TLNs of a NS depend on the underlying EOS of nuclear matter. In particular, for a given mass, different prescriptions for the EOS give rise to different radii and tidal deformabilities. However, it was found that the moment of inertia, the spin-induced quadrupole moment, and the electric quadrupolar tidal deformability are related to each other by nearly EOS-independent relations known as the I-Love-Q relations. The latter are approximately universal at the level of $\sim 1 \%$, and hold even for strange quark stars~\cite{Yagi:2013bca,Yagi:2013awa,Yagi:2016ejg}.

Similar approximately universal relations for NSs hold also between their quadrupolar electric and magnetic tidal deformabilities~\cite{Yagi:2013sva}, with an accuracy of about $(1 \divisionsymbol 10) \%$. In particular, the fitting function 
\begin{align}
\log|\bar{\sigma}_2| &= a_i + b_i \log\bar{\lambda}_2 + c_i (\log\bar{\lambda}_2)^2 \nonumber \\
& + d_i (\log\bar{\lambda}_2)^3 + e_i (\log\bar{\lambda}_2)^4 +  f_i (\log\bar{\lambda}_2)^5 \,,
\end{align}
with fitting coefficients $a_i = -2.03$, $b_i = 0.487$, $c_i = 9.69 \cdot 10^{-3}$, $d_i = 1.03 \cdot 10^{-3}$, $e_i = -9.37 \cdot 10^{-5}$, $f_i = 2.24 \cdot 10^{-6}$ is a good approximation for realistic NSs described by an irrotational fluid~\cite{Jimenez-Forteza:2018buh}. This fit is shown by the red solid line in the right panel of Fig.~\ref{universalityNS}. As expected, as one increases the compactness, NSs approach the BH limit with vanishing tidal deformability. 

Figure~\ref{universalityNS} also shows the corresponding approximate universality relations for FSSs. In the left panel, one can appreciate that for FSSs the $\bar\sigma_2-\bar\lambda_2$ relations are less universal than for NSs, since the curves display $\mathcal{O}(1)$ corrections in the small compactness regime for different choices of the model parameters, while they are less affected in the large compactness region. The emergence of approximate universal relations can be understood by inspecting Eqs.~\eqref{eq:k2E} and \eqref{eq:k2B}. They both depend on two dimensionless parameters, $C$ and $y$. Recall that both mass and radius scale approximately as $\sim 1/\Lambda^\alpha$, where $\alpha = 1$ for $\zeta \neq 0.5$, and $\alpha = 2$ for $\zeta = 0.5$ (see Fig.~5 of Ref.~\cite{DelGrosso:2024wmy}), while the dependence on $\eta,\,\zeta$ is weak. From the definitions of $C$ and $y$, we see that they are both left unchanged under such a rescaling. Thus, the TLNs will also be mildly dependent on the fundamental parameters of the theory. This dependence may become weaker and weaker for small values of $\Lambda$, which are however numerically challenging to achieve. Indeed, as shown in Fig.~5 of Ref.~\cite{DelGrosso:2024wmy}, the universal scaling of the mass-radius diagram becomes more and more accurate for small $\Lambda$, and one expects to find a similar behavior for the TLNs.

In the right panel of Fig.~\ref{universalityNS}, we compare the $\bar\sigma_2-\bar\lambda_2$ relations for FSSs with those for realistic NSs. We find that the FSS relation differs from the NS curve, highlighting the different nature of these objects.
By comparing the left and right panels of Fig.~\ref{universalityNS} we see that, although the TLNs of a FSS are less universal than those of a NS, the difference in the quasi-universal relations among these classes of objects are much larger then their individual spread. Thus, a sufficiently accurate measurement can tell the two relations apart.

The internal structure of FSSs allows for a comparison with the bag model described by the EOS~\cite{Urbano:2018nrs}
\begin{equation}
  W = W_0 + P/\omega \,,
\end{equation}
which corresponds to the stiffest possible EOS, as the speed of sound $c_s = \sqrt{\omega}$ takes the maximal value throughout the object. By renormalizing the pressure and density to the central value $W_0$, one can show that this model allows for a maximal compactness of~\cite{Urbano:2018nrs}
\begin{equation}
C_\text{\tiny Bag} \leq \frac{4}{9} \frac{\omega (4.18 + \omega)}{0.77 + 4.69 \omega + \omega^2}\,,
\end{equation}
quantifying the distance from the Buchdahl bound (obtained in the incompressible fluid limit as $\omega\to\infty$). Notice that a nonzero value of the central density $W_0$ is necessary to ensure a finite radius.

It is therefore interesting to compare the relativistic TLNs of a FSS with those of the bag model assuming large internal pressure and sound speed $\omega = 1/3$, which has a maximum compactness of about $C_\text{\tiny Bag} \lesssim 0.4$. This comparison is motivated by the fact that, within its interior, the FSS is dominated by relativistic fermions, and the scalar field plays a negligible role.
Following this comparison, one can solve the corresponding perturbation equations for the bag model to derive both the electric and magnetic TLNs. Notice that the presence of a discontinuity in the energy density at the object's surface induces a change in the boundary condition for the computation of the electric TLNs, which corresponds to shifting the parameter $y$ by $- 4 \pi W_0 R^3/M$~\cite{Hinderer:2009ca}. The corresponding universality relations are shown by the water green curve in the right panel of Fig.~\ref{universalityNS}. Similarly to the findings of Ref.~\cite{Yagi:2013bca}, they are not only in agreement with the standard fit for NSs with ordinary EOS, but the final result is also very mildly dependent on the sound speed $\omega$. This suggests that the bag model does not capture the deformability properties of FSSs: the presence of the scalar field makes a sizeable contribution to the overall energy, and dominates the object's outer region. 

The right panel of Fig.~\ref{universalityNS} also shows the universality relations for different families of boson stars~\cite{Cardoso:2017cfl}. These complex bosonic self-gravitating configurations experience different tidal deformations compared to the other objects, which strongly depend on the properties of the scalar field potential.

Finally, we notice the presence of a hierarchy between the tidal deformabilities of each compact object in the configurations with maximal compactness  (identified by the lower edges of each curve in Fig.~\ref{universalityNS}). In particular, the bag model admits a higher (lower) electric (magnetic) TLN compared to FSSs. Among the various families of boson stars, the minimal model with no scalar interaction displays the largest tidal deformability, while solitonic boson stars are less (more) deformable than massive ones in the electric (magnetic) sector.
Accurate tidal deformability measurements can be used to identify different families of compact objects. 

As discussed above, we focused on the most interesting case of an irrotational fluid. For a static fluid, the magnetic TLNs of a NS have the opposite sign and are quantitatively different. They display an approximately universal relation different from irrotational fluid NSs~\cite{Delsate:2015wia}, as shown by the red dashed line of Fig.~\ref{universalityNS} (obtained from the corresponding fit of Ref.~\cite{JimenezForteza:2018rwr}). 
Interestingly, next-generation ground-based GW detectors~\cite{Kalogera:2021bya}, such as the Einstein Telescope and Cosmic Explorer, should allow us to measure $\sigma_2$ and $\lambda_2$ with sufficient precision to distinguish the irrotational-fluid case from the static-fluid case~\cite{JimenezForteza:2018rwr}. Since the difference between the NS curve and the FSS curve shown in Fig.~\ref{universalityNS} is even larger, future detectors should be able to distinguish FSSs from NSs based on tidal deformability measurements.

Furthermore, from Figs.~\ref{FSSTLN} and \ref{universalityNS} we see that FSSs have lower deformability than other ECOs for certain model parameters, so it could be difficult to tell them apart from black holes through the measurement of tidal effects in gravitational waveforms~\cite{Crescimbeni:2024cwh}. It would be interesting to quantify these expectations with more detailed parameter estimation calculations. 

\subsection{Tidal disruption}

Let us consider a massive, nonspinning central object whose mass and radius are denoted by $M_B$ and $R_B$, respectively. It is interesting to ask whether a FSS with radius $R$ and mass $M$, orbiting around the central object, can be tidally disrupted, assuming $M \ll M_B$. To this end, we need to estimate the Roche radius of the system
\begin{equation}\label{eq:roche}
R_\text{\tiny Roche} \sim \gamma R_{B} \left( \frac{W_{B}}{W} \right)^{1/3} \, ,
\end{equation}
where $W_{B}$, $W$ are the densities of the central object and FSS, respectively, and the numerical coefficient $\gamma$ takes values ranging from 1.26 for rigid bodies to 2.44 for fluid bodies~\cite{1983bhwd.book.....S}.

Whenever $\zeta > 1/2$ (asymmetric vacua regime), hydrostatic equilibrium imposes (introducing $q = \sqrt{\mu v_{\rm F}}$)
\begin{equation}
    \omega_{\rm F} \sim \frac{m_f}{\eta} = q \, ,
\end{equation}
as long as we are above the minimal configuration in the mass-radius diagram, corresponding to the minimum value for $\omega_{\rm F}$~\cite{DelGrosso:2024wmy}
(we do not take into account configurations along the mass-radius diagram under the minimal one because the initial displacement of the scalar field becomes $\mathcal{O}(1)$, and thus true solitonic configurations describing false vacuum pockets are not allowed: see Ref.~\cite{DelGrosso:2024wmy}). Thus, the density $W$ is estimated as the central fermion energy density $W_c \sim k_{\rm F}^4 \sim \omega^4_{\rm F} \sim q^4$.

The physically most compelling scenario is when the tidal disruption happens before the merger phase. Imposing $R_\text{\tiny Roche} > R_\text{\tiny ISCO} = 6 \,G M_B$ in Eq.~\eqref{eq:roche}, we find the condition (ignoring $\mathcal{O}(1)$ factors)
\begin{equation}
    q \lesssim \Big(\frac{m_p}{M_B}\Big)^{1/2} m_p \approx 1.3\,{\rm GeV} \Big(\frac{M_\odot}{M_B}\Big)^{1/2}\,.
\end{equation}
For an astrophysical solar-like object with $M_B \sim 1 \,M_\odot$, the latter condition implies $q \lesssim 1\, {\rm GeV}$. Higgs false vacuum balls and dark soliton stars~\cite{DelGrosso:2024wmy} require $q \gtrsim 10^2 \, {\rm GeV}$, and therefore do not get tidally disrupted before the merger. Instead, neutron soliton stars (also known as quark nuggets) correspond to $q  \approx 0.2 \, {\rm GeV}$, potentially allowing for tidal disruption of these nontopological solitons, in a range of masses and radii starting from the (noncompact) minimal configuration
\begin{align}\label{}
 M_{\rm min} \approx 10^{-19} \,M_\odot, \qquad 
   R_{\rm min} \approx 1\, {\rm cm}\,,
\end{align}
up to the critical one 
\begin{align}\label{benchmark_values_NS}
   M_c \approx 2 \,M_\odot, \qquad 
   R_c \approx 10\, {\rm km}\,,
\end{align}
with a compactness of $G M_c/R_c\sim 0.27$, which is
slightly larger than that of an ordinary NS. In the latter case, the mass of the FSS becomes comparable to the mass of the central object, and thus the previous estimates should be taken with a grain of salt, although they should provide the correct order of magnitude.

In a tidal disruption scenario, the quarks released in the disruption event would produce jets of hadrons and photons~\cite{Witten:1984rs, 1985ICRC....8..290A, Bai:2018dxf}. Assuming that the amount of energy emitted during a collision is of the order of $ f_\text{\tiny rad} (G M_B M/R_\text{\tiny Roche})$ and considering an efficiency factor $f_\text{\tiny rad}$ for the energy going into visible Standard Model radiation, the corresponding power output in one orbital period is given by
\begin{align}
\mathcal{P} \simeq 10^{21} L_\odot f_\text{\tiny rad} \left(\frac{M}{0.1 M_\odot} \right) \left(\frac{M_B}{M_\odot} \right)^{2/3} \left(\frac{q}{0.2 \, {\rm GeV}} \right)^{10/3}\,,
\end{align}
where $L_\odot = 3.8 \cdot 10^{26} \, {\rm W}$ is the luminosity of the Sun. This power is comparable to the one emitted by superradiance (see Ref.~\cite{Brito:2015oca} for a discussion).

To assess whether a telescope on Earth could detect this radiation, we assume an angular resolution of $\delta \Omega = 1^\circ \times 1^\circ =  (\pi/180)^2  \, {\rm sr}$ and that the binary is situated at a distance $d \simeq 1\,{\rm Gpc}$. Then, the frequency-weighted spectral density is estimated to be
\begin{align}
\nu I_\nu = \frac{\mathcal{P}}{d^2 \delta \Omega} & \simeq  10^{-9} \frac{\rm W}{\rm m^2 sr}   \lp \frac{f_\text{\tiny rad}}{10^{-10}} \rp \lp \frac{d}{\rm Gpc} \rp^{-2}
\nonumber \\
& \times \left(\frac{M_B}{M_\odot} \right)^{2/3} \left(\frac{M}{0.1 M_\odot} \right)  \left(\frac{q}{0.2 \, {\rm GeV}} \right)^{10/3}\,.
\end{align}
For comparison, the observed cosmic backgrounds of X-rays and gamma rays range from $10^{-10} {\rm W/m^2 sr}$ at energies around $10 \, {\rm keV}$ to $10^{-13} {\rm W/m^2 sr}$ around $10 \, {\rm GeV}$~\cite{Hill:2018trh}, implying that the tidal disruption of these quark nuggets may produce detectable photons for an efficiency factor as small as $10^{-10}$. If the central object is a BH, a sizeable fraction of the emitted matter could be accreted by the central object, giving rise to a subsequent afterglow.

\section{Conclusions}
\label{sec:conclusions}
In this paper we studied the deformability properties of FSSs, solutions of general relativity in which a real scalar field is coupled to a fermionic field by a Yukawa coupling. The coupling generates an effective mass for the fermions as the scalar field transitions from a false vacuum to a true vacuum configuration.
The structure of vacua in the scalar field potential determines the nature of the corresponding compact objects, whose mass-radius curves exhibit different phenomenology for different model parameters.

In this work we have studied both electric- and magnetic-type perturbations of a FSS background. By perturbing both the scalar and fermionic sectors (and assuming the latter to be described by the Thomas-Fermi approximation also in the perturbed configuration), we have derived the corresponding perturbation equations and solved them to obtain the conservative and irrotational TLNs.
These are found to depend on the model parameters, especially on the scalar field vacuum expectation value. As in the case of NSs, the magnetic TLNs are generally smaller than the electric TLNs. 

Using the tidal deformabilities computed in this way, we have investigated the existence of approximately universal relations between the Love numbers in both parity sectors, showing a mild dependence on the model parameters and therefore a solid prediction for these relations (even though they are less universal than the ones for NSs). We then compared the quasi-universal relations for FSSs with those found for other compact objects, such as NSs and boson stars, showing that the universality relations corresponding to different classes of compact objects are significantly different. This feature could be used as a novel probe to tell apart various classes of compact objects using tidal deformability measurements with next-generation detectors. 

The characteristic values of the TLNs of FSSs (for model parameters corresponding to objects in the solar mass range) imply that these quantities may be measurable by future gravitational wave interferometers, such as the Einstein Telescope and Cosmic Explorer. In particular, the results of Ref.~\cite{JimenezForteza:2018rwr} imply that these instruments could measure both the  electric and magnetic Love numbers with an accuracy of a few percent, potentially allowing us to distinguish FSSs from other compact objects in the solar mass range, such as ordinary NSs or black holes~\cite{Crescimbeni:2024cwh}. 

Finally, we discussed the possible disruption of a FSS in a binary system with another compact object, such as a black hole, and derived a bound on the vacuum expectation values that may allow for a tidal disruption event before the ISCO frequency is reached. Such tidal disruption events could happen for nontopological quark nuggets, which could release jets of hadrons and photons during the event. If this process occurs in nature, it would provide a significant contribution to the observed cosmic backgrounds of X-rays and gamma rays, and even to single resolvable events.

This work is an initial step towards a full investigation of the tidal interactions and tidal deformabilities of ECOs, and it can be improved in various directions. It would be interesting to perform a more detailed study to understand how tidal effects could be used to distinguish different classes of ECOs. It is important to generalize the computation beyond the simplifying assumption of spherical symmetry, in order to assess if the I-Love-Q relations are valid also for rotating FSSs. Finally, given the intrinsic time dependence in the evolution of a binary system, it would be interesting to investigate the FSS dissipative coefficients and frequency-dependent TLNs, as recently discussed in Refs.~\cite{Nair:2022xfm, Chakraborty:2023zed} for Kerr-like compact objects.
We leave these studies for future work.

\section*{Acknowledgments}
We thank Konstantinos Kritos and Kent Yagi for useful discussions.
E.B. is supported by NSF Grants No. AST-2006538, PHY-2207502, PHY-090003 and PHY-20043, by NASA Grants No. 20-LPS20-0011 and 21-ATP21-0010, by the John Templeton Foundation Grant 62840, by the Simons Foundation, and by the Italian Ministry of Foreign Affairs and International Cooperation Grant No. PGR01167.
V.DL. is supported by funds provided by the Center for Particle Cosmology at the University of Pennsylvania. 
L.DG. acknowledges the Johns Hopkins University for hospitality during the completion of this project and H2020-MSCA-RISE-2020 GRU (Grant agreement ID: 101007855) for financial support.
P.P. is partially supported by the MUR PRIN Grant 2020KR4KN2 ``String Theory as a bridge between Gauge Theories and Quantum Gravity'' and by the MUR FARE programme (GW-NEXT, CUP:~B84I20000100001).

\appendix

\section{Fermionic perturbations}\label{appendixA}

In the axial sector, since $\phi_1 \equiv 0$, the $\theta \theta $ plus the $\varphi \varphi$ component of the Einstein equations gives simply $\delta k_{\rm F} = 0$. Thus, $\delta P = \delta W = \delta S = 0$. Conversely, in the polar sector, the same combination of the Einstein equations gives
\begin{align}\label{eq:deltaP}
P_1 = \frac{1}{2} H_0 (\Bar{P} + \Bar{W}) + U_1 +  \Big(f \Bar{S} -\frac{\partial U}{\partial \phi}\Big|_{\Bar{\phi}}\Big) \phi_1 \,.
\end{align}
Following the procedure described in Sec.~\ref{sec:fermionic_perturbations}, we express $k_{\rm F} = (k_{\rm F})_1(\rho) Y(\theta, \varphi)$ (omitting the multipolar indices) as
\begin{widetext}
\centering
\begin{align}\label{eq:deltakF}
(k_{\rm F})_1 & = \frac{1}{4 \Bar{k}_\F^4} \bigg[-\phi_1 (3 \sqrt{\Bar{k}_\F^2+ m_\eff^2} (4 \pi^2
   \partial U / \partial \phi\big|_{\Bar{\phi}}+f m_\eff^3 \Big(\log
   \Bigg(1-\frac{\Bar{k}_\F}{\sqrt{\Bar{k}_\F^2+m_\eff^2}}\Bigg) \nonumber \\
   &-\log
  \Bigg(\frac{\Bar{k}_\F}{\sqrt{\Bar{k}_\F^2+m_\eff^2}}+1\Bigg)\Big)-4 \pi^2 f
   \Bar{S})+2 f \Bar{k}_\F^3
   m_\eff+6 f \Bar{k}_\F m_\eff^3) \nonumber\\
   &+6 \pi^2 H_0\sqrt{\Bar{k}_\F^2+m_\eff^2} (\Bar{P}+\Bar{W})+12 \pi^2 U_1\sqrt{\Bar{k}_\F^2+m_\eff^2}\, \bigg] \,,
\end{align}
\end{widetext}
where $m_\eff = m_f - f \Bar{\phi}$.

Expanding Eqs.~\eqref{fermion_energy},~\eqref{fermion_density} and substituting Eq.~\eqref{eq:deltaP} and Eq.~\eqref{eq:deltakF}, we obtain $W_1, S_1$ in terms of the background quantities $m_{\rm eff}$ and $\Bar{k}_{\rm F}$, as well as metric and scalar perturbations, as follows:
\begin{widetext}
\centering
\begin{align}
W_1 & = \frac{1}{4 \pi^2 \Bar{k}_\F^2 \sqrt{\Bar{k}_\F^2+m_\eff^2}} \bigg[3 m_\eff^2 \sqrt{\Bar{k}_\F^2+m_\eff^2} (\phi_1
   (-4 \pi^2 \partial U / \partial \phi\big|_{\Bar{\phi}}+f m_\eff^3 (\log (\frac{\Bar{k}_\F}{\sqrt{\Bar{k}_\F^2+m_\eff^2}}+1) \nonumber \\
   &-\log (1-\frac{\Bar{k}_\F}{\sqrt{\Bar{k}_\F^2+m_\eff^2}}))+4 \pi^2 f \Bar{S})+4 \pi^2 U_1)+\Bar{k}_\F^2
   \sqrt{\Bar{k}_\F^2+m_\eff^2}
   (\phi_1(-12 \pi^2
   \partial U / \partial \phi\big|_{\Bar{\phi}} \nonumber\\
   &+f m_\eff^3 (-3 \log
   (1-\frac{\Bar{k}_\F}{\sqrt{\Bar{k}_\F^2+m_\eff^2}})+3 \log
   (\frac{\Bar{k}_\F}{\sqrt{\Bar{k}_\F^2+m_\eff^2}}+1)+2 \tanh^{-1}\Bigg(\frac{\Bar{k}_\F}{\sqrt{\Bar{k}_\F^2+m_\eff^2}}\Bigg))+12 \pi^2 f
   \Bar{S}) \nonumber\\
   &+12 \pi ^2 U_1)-4
   f \Bar{k}_\F^5 m_\eff \phi_1-10 f \Bar{k}_\F^3 m_\eff^3
   \phi_1-6 f \Bar{k}_\F
   m_\eff^5 \phi_1+6 \pi^2
  H_0(\Bar{k}_\F^2+m_\eff^2)^{3/2} (\Bar{P} + \Bar{W})\bigg]\,, \\
  S_1 & = \frac{1}{4 \pi ^2 \Bar{k}_\F^2 \sqrt{\Bar{k}_\F^2+m_\eff^2}} \bigg[12 \pi ^2 m_\eff\sqrt{\Bar{k}_\F^2+m_\eff^2} (\phi_1 (f
   \Bar{S}-\partial U / \partial \phi\big|_{\Bar{\phi}})+U_1)-3
   fm_\eff^4 \phi_1 \nonumber\\
   & \left(\sqrt{\Bar{k}_\F^2+m_\eff^2}
   \left(\log \left(1-\frac{\Bar{k}_\F}{\sqrt{\Bar{k}_\F^2+m_\eff^2}}\right)-\log \left(\frac{\Bar{k}_\F}{\sqrt{\Bar{k}_\F^2+m_\eff^2}}+1\right)\right)+2\Bar{k}_\F\right)-2 f \Bar{k}_\F^2m_\eff^2 \phi_1 \nonumber\\
   & \left(4 \Bar{k}_\F-3 \sqrt{\Bar{k}_\F^2+m_\eff^2} \tanh
   ^{-1}\left(\frac{\Bar{k}_\F}{\sqrt{\Bar{k}_\F^2+m_\eff^2}}\right)\right)-2 f
   \Bar{k}_\F^5 \phi_1+6 \pi^2
   H_0 m_\eff \sqrt{\Bar{k}_\F^2+m_\eff^2} (\bar{P}+\bar{W}))\bigg]\,.
\end{align}
\end{widetext}

\bibliography{refs}

\end{document}